\documentclass[preprint,review,12pt]{elsarticle}
\usepackage{marginnote}

\usepackage{lineno,hyperref}
\modulolinenumbers[5]

\usepackage{hyphenat}

\usepackage{graphicx}
\usepackage{color}

\usepackage{amssymb}
\usepackage{amsmath}
\usepackage{gensymb}
\usepackage{textcomp}
\usepackage{upgreek}
\usepackage{gensymb}
\usepackage[version=4]{mhchem}
\usepackage{booktabs}
\usepackage{adjustbox}
\usepackage{rotating}

\journal{Journal of Physics and Chemistry of Solids}

\begin{document}
\sloppy
\begin{frontmatter}

\title{Chemical-state analyses of Ni, Zn, and W ions in \ce{NiWO4}-\ce{ZnWO4} solid solutions by X-ray photoelectron spectroscopy}

\author[myaddress]{G. Bakradze
\corref{mycorrespondingauthor}}
\cortext[mycorrespondingauthor]{Corresponding author}
\ead{georgijs.bakradze@cfi.lu.lv}
\author[myaddress]{A. Kalinko}
\author[myaddress]{A. Kuzmin}

\address[myaddress]{Institute of Solid State Physics, University of Latvia, Kengaraga Street 8, Riga, LV-1063, Latvia}

\begin{abstract}
The chemical states of Ni, Zn, and W in microcrystalline \ce{NiWO4}-\ce{ZnWO4} solid solutions were studied by X-ray photoelectron spectroscopy. The recorded spectra of the Ni 2p, Zn 2p, and W 4f photoelectron lines and Ni $\text{L}_2\text{M}_{23}\text{M}_{45}$, Zn $\text{L}_3\text{M}_{45}\text{M}_{45}$, and W $\text{N}_4\text{N}_{67}\text{N}_{7}$ Auger-transition lines show pronounced changes with increasing Zn concentration. The positions of the resolved photoelectron and Auger-transition lines were combined to construct so-called chemical-state plots (Wagner or Auger-parameter plots) for metal ions in solid solutions. With increasing Zn concentration, the Auger parameter increases for Ni and decreases for W, thus evidencing a lowering and an increase of the electronic polarizability around core-ionized Ni and W ions, respectively. At the same time, the character of Zn--O bonds and the local structure around Zn ions do not change. It is concluded that the dilution of \ce{NiWO4} with Zn ions is accompanied by an increase of the Ni--O bond ionicity and an increase of the W--O bond covalency. These changes are attributed to the charge redistribution among \ce{[NiO6]} and \ce{[WO6]} structural units. We show that a careful in-depth analysis of XPS data obtained with a laboratory-based X-ray photoelectron spectroscopy system can give chemically sensitive, qualitative information on the changes in the first coordination spheres of each metal ion. This information is otherwise accessible only by synchrotron-based techniques (such as X-ray absorption spectroscopy).
\end{abstract}


\end{frontmatter}

\section{Introduction}
Tungstates have technologically important applications as scintillators \cite{Nikl2008}, laser host materials \cite{Pask2003}, photocatalysts \cite{H2021121892}, and heterogeneous catalysts \cite{Yan2019}. Many properties of tungstates can be further controlled by doping or by making solid solutions with other tungstates \cite{Schofield1993, Yu2003, Isupov2005b, Kaczmarek2013, Dey2014, Kuzmin2016}. The doping approach is of particular interest because of a wide range of possible chemical compositions.

Both \ce{ZnWO4} and \ce{NiWO4} crystallize in the wolframite structure (space group $P2/c$) with two formula units per unit cell \cite{Filipenko1968, WEITZEL1970}. The structure comprises infinite zigzag chains running in the [001] direction and formed entirely of either edge-sharing \ce{[$A$O6]} octahedra or edge-sharing \ce{[WO6]} octahedra. Each chain of \ce{[$A$O6]} octahedra is corner-linked to four chains of \ce{[WO6]} octahedra, and vice versa, leaving open channels in the [001] direction. From X-ray diffraction and X-ray absorption spectroscopy studies, it is commonly accepted that the structure of \ce{$A$_c$B$_{1-c}WO4} binary solid solutions  of isomorphous wolframite-type tungstates \ce{$A$WO4} and \ce{$B$WO4} consists of independent \ce{[$A$O6]}, \ce{[$B$O6]}, and \ce{[WO6]} octahedra. However, multiple articles showed that on a microscopic level the local structure in these (or similar) materials can be quite complex \cite{Schofield1993, Kuzmin2016}. The formation of solid solutions also changes the environment around $B$ cations and increases the unit cell volume in proportion to $A$ concentration (for \ce{Zn_cNi_{1-c}WO4}, see \cite{Kalinko2011a, Bakradze2020, BAKRADZE2021}; the ionic radii are  0.72 $\AA$ and 0.69 $\AA$ for Zn and Ni, respectively \cite{lide2007}). In isomorphous systems, these changes occur continuously and variably in a controlled manner as a function of $A$ concentration in the whole solid-solution range \cite{Karmakar2020}.

One of the tools that can probe the local environment of an element is X-ray photoelectron spectroscopy (XPS). XPS is a powerful surface analytical technique to characterize the (different) chemical bonding states of elements in near-surface regions (i.e., up to depths of about 10~nm) of a solid compound. For example, different oxidation states of an element in a compound can often be distinguished on the basis of the chemical shifts of the core-level X-ray-induced photoelectron lines with respect to some well-defined reference state (e.g., the metallic state or free atom in the gas phase) \cite{MORETTI1998, MORETTI2013}. The local chemical state of an element in a solid can be particularly effectively assessed by XPS on the basis of the so-called modified Auger parameter (AP), as first introduced by Wagner \cite{WAGNER1988}. The modified AP (for simplicity we further refer to it simply as the AP) of an element in a compound, $\alpha$, is defined as the sum of the binding energy (BE) of the most prominent and sharpest core-level photoelectron line ($c_0$) and the kinetic energy (KE) of the most prominent and sharpest core-level-like Auger-transition line ($c_1c_2c_3:\,^{2S+1}L_J$):

\begin{equation}
\alpha = E_\text{b}(c_0) + E_\text{k}(c_1c_2c_3:\,^{2S+1}L_J).
\label{eq:eq1}
\end{equation}

The AP provides a direct measure of the electronic polarizability of the chemical environment around the core-ionized atom and is, therefore, sensitive to structural changes in the nearest coordination sphere of the element considered. The AP $\alpha(c_0; c_1c_2c_3:\,^{2S+1}L_J)$ may be displayed in a diagram called a \textit{chemical-state plot} (also known as a Wagner plot or an AP plot), which is of considerable analytical utility \cite{MORETTI2013, Bakradze2011}.

The value of the AP is independent of the selected energy reference level (i.e., the position of the Fermi level in the band gap) and, thus, is unaffected by energy shifts of the measured photoelectron and Auger-transition lines due to charging (as encountered in the XPS analysis of insulating compounds). Against this background and recognizing the high surface sensitivity offered by XPS, the AP provides a unique tool for the direct experimental assessment of the changes in the local chemical state of elements in complex oxides as a function of concentration. The AP shift between two states can be expressed as
\begin{equation}
\Delta \alpha  \approx 2 \Delta R^\text{ea},
\label{eq:eq2}
\end{equation}
where $R^\text{ea}$ is the extra-atomic relaxation energy. Furthermore, Wagner \cite{WAGNER1982, WAGNER1988} showed that the shifts in core-level BE, $\Delta E_\text{b}$, and the shifts in Auger-transition KE, $\Delta E_\text{k}$, can be expressed, respectively, as
\begin{equation}
\Delta E_\text{b} = - \Delta \epsilon - \Delta R^\text{ea},
\label{eq:eq3}
\end{equation}
\begin{equation}
\Delta E_\text{k} = \Delta \epsilon + 3 \Delta R^\text{ea},
\label{eq:eq4}
\end{equation}
where $\Delta \epsilon$ is the so-called initial-state contribution.

While the use of chemical-state plots to study a complex oxide is not very widespread, such an approach can give remarkable insights into the chemical environment of the elements \cite{Bakradze2011, Christensen2013, Moretti2020}. In this study, the chemical-state analyses of metal ions in \ce{Zn_cNi_{1-c}WO4} solid-solution powders were performed with the APs of Ni, Zn, and W to reveal the chemical-state evolution of these elements as a function of composition.

\section{Experimental and data analysis}\label{s:exper}
\ce{Zn_cNi_{1-c}WO4} solid solutions were obtained by coprecipitation at room temperature (20~$\celsius$) and pH~8 with \ce{ZnSO4.7H2O}, \ce{Ni(NO3)2.6H2O}, and \ce{Na2WO4.2H2O} as educts (see \cite{Kalinko2011a, Bakradze2020} for details). First, the salt solutions were dissolved in double-distilled water. Next, the salt solutions containing metal cations were mixed in stoichiometric ratios and added to the sodium tungstate solution to obtain precipitates of \ce{Zn_cNi_{1-c}WO4} solid solutions. The products precipitated immediately upon mixing of the aqueous solutions. After completion of the reaction, the precipitates were filtered off, washed several times with double-distilled water, and dried in air for 12~h at 80~$\celsius$. Microcrystalline powders were produced by our annealing the precipitates in an ambient atmosphere for 4~h at 800~$\celsius$. The phase composition of the samples was checked by X-ray powder diffraction. To produce pristine surfaces for analysis, the powders were ground in 2-propanol (analytical purity) with use of an agate mortar, and then the suspension was quickly transferred to a stainless steel sample holder; the drying was completed under a vacuum in the entry lock of the XPS system.

\begin{figure}
	\centering
	\includegraphics[width=0.4\textwidth]{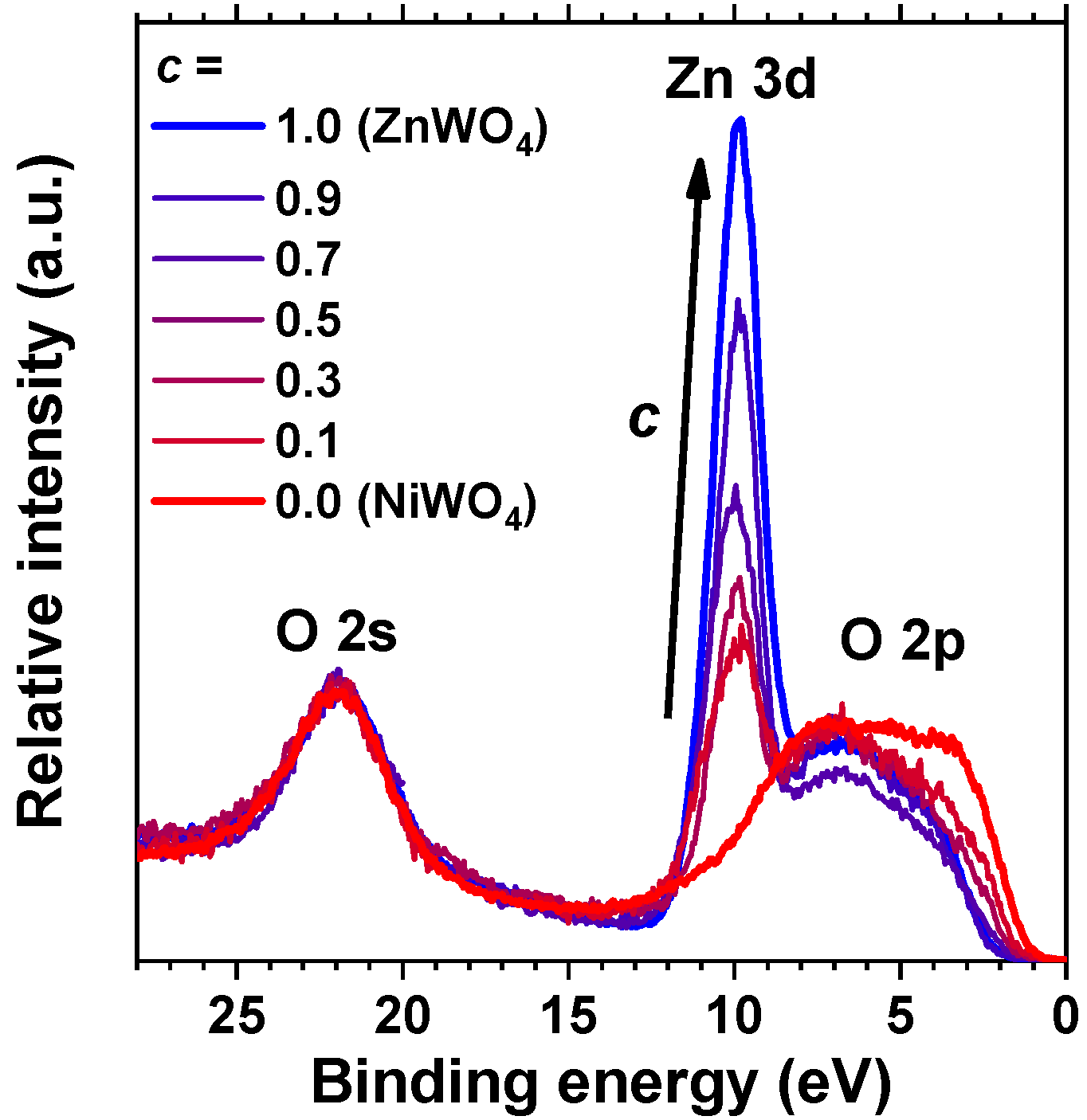}
	\caption{X-ray photoelectron spectroscopy valence band spectra as recorded from microcrystalline \ce{Zn_cNi_{1-c}WO4} solid solutions. The positions of the O 2s and Zn 3d photoelectron peaks are indicated. The arrow indicates the direction of increasing zinc concentration. The spectra have been scaled and shifted for alignment with the O 2s peak.}
	\label{fig1}
\end{figure}

XPS analyses of the powders were conducted with a Thermo Scientific ESCALAB Xi+ system using monochromatic Al K$_\alpha$ radiation ($h\nu = 1486.68$ eV, 14.8~kV, 175~W) for all samples at room temperature; the BE scale was calibrated according to \cite{Seah2001}. The resolution function of the instrument was found to have a width of 0.37 eV with use of  the Ag Fermi edge. Spectra were collected under zero angle to the sample normal, and averaged over three to five different spots on the samples. The base pressure in the analysis chamber was lower than $3 \times 10^{-8}$ Pa. To control charging of the samples in all experiments, a dual electron-ion flood gun (2~V, 100~ $\upmu$A) was used; during its operation the working pressure was around $7\times 10^{-5}$ Pa. To determine the position of the C 1s peak, a single peak (Gaussian-Lorentzian product function with 70\% Gaussian contribution)---ascribed to alkyl-type carbon (C--C, C--H)---was fit to the main peak of the C 1s spectrum for adventitious carbon. On the high-BE side, the main peak was always accompanied by a second peak shifted by about +1.5~eV. The higher-BE peak is ascribed to an alcohol (C--OH) functionality. The BEs of the recorded spectra were charge-corrected by setting the alkyl C 1s peak BE position to 285.0~eV, which typically meant the recorded spectra had to be shifted by approximately +1.2~eV. The integral area of the C 1s peak never exceeded 5\% of that of the most intense photoelectron peak in the survey.

Inhomogeneous charging can broaden peaks recorded by XPS, but as can be seen from Fig.~\ref{fig1}, the width and the shape of the O 2s photoelectron peak stay the same from sample to sample; thus, this proves that the charging does not change considerably in the series. The X-ray photoelectron spectra of the Ni 2p, Zn 2p, and W 4f regions were recorded over the BE ranges from 850~eV to 890~eV, from 1010~eV to 1050~eV, and from 33~eV to 44~eV, respectively; the most prominent Auger-transition lines---Ni $\mathrm{L}_2\mathrm{M}_{23}\mathrm{M}_{45}$, Zn $\mathrm{L}_3\mathrm{M}_{45}\mathrm{M}_{45}$, and W $\mathrm{N}_4\mathrm{N}_{67}\mathrm{N}_{7}$---were recorded over the BE ranges from 630~eV to 660~eV, from 485~eV to 510~eV, and from 1300~eV to 1340~eV, respectively. For simplicity, we refer to these Auger transitions as Ni LMM, Zn LMM, and W NNN, respectively. All spectra were recorded with a step size of 0.05~eV at a constant pass energy of 20~eV with a dwell time of 50~ms. The valence band (VB) spectra were recorded as part of the W 4f spectrum. The measured spectra were all corrected for the electron-KE-dependent transmission of the analyzer by use of the corresponding correction factor as provided by the manufacturer. Furthermore, the lower-BE side of the thus-corrected X-ray photoelectron spectra was set to zero (background) intensity by subtraction of a constant background, the value of which was taken to be equal to the averaged minimum intensity at the lower-BE side of the corresponding peak envelope. In \ce{NiWO4}, the Ni 2p doublet possesses an intense, complex multiplet splitting and poses unique difficulties for chemical-state analysis \cite{BIESINGER20112717}. Following the approach described in \cite{GROSVENOR20061771}, we used a dedicated procedure for compounds containing high-spin Ni cations using the free-ion Gupta-Sen multiplets and shake-up-related satellites to fit the Ni 2p$_{3/2}$ envelope. Next, the BE positions of the predominant W 4f$_{7/2}$ and Zn 2p$_{3/2}$ peaks were resolved by a standard fitting procedure using one peak and two peaks of the Gaussian-Lorentzian product function (70\% Gaussian contribution), respectively. During the fitting procedure, for all photoelectron peaks the background from inelastically scattered electrons was subtracted with use of the Shirley-Proctor-Sherwood background subtraction method by means of the active approach \cite{HerreraGomez2014}. The positions of the main peaks and the Ni LMM, Zn LMM, and W NNN Auger-transition lines were straightforwardly determined from the position of the most intense peak in the respective Auger-transition region, which was fit by either two or three Gaussian-Lorentzian product functions (70\% Gaussian contribution) with use of an active linear background \cite{Sherwood2019}. The numerical errors in the fitted BEs or KEs were estimated as described in \cite{Cumpson1992}. The O 1s and O KLL regions were not considered, because it was not possible to reliably determine all contributions to the spectra (there are: several chemically different O positions in the structure of pure wolframites; adsorbed O-containing species; surface hydroxides; moreover, the O KLL region overlaps with the Ni 2s peak).

\begin{figure}
	\centering
	\includegraphics[width=0.4\textwidth]{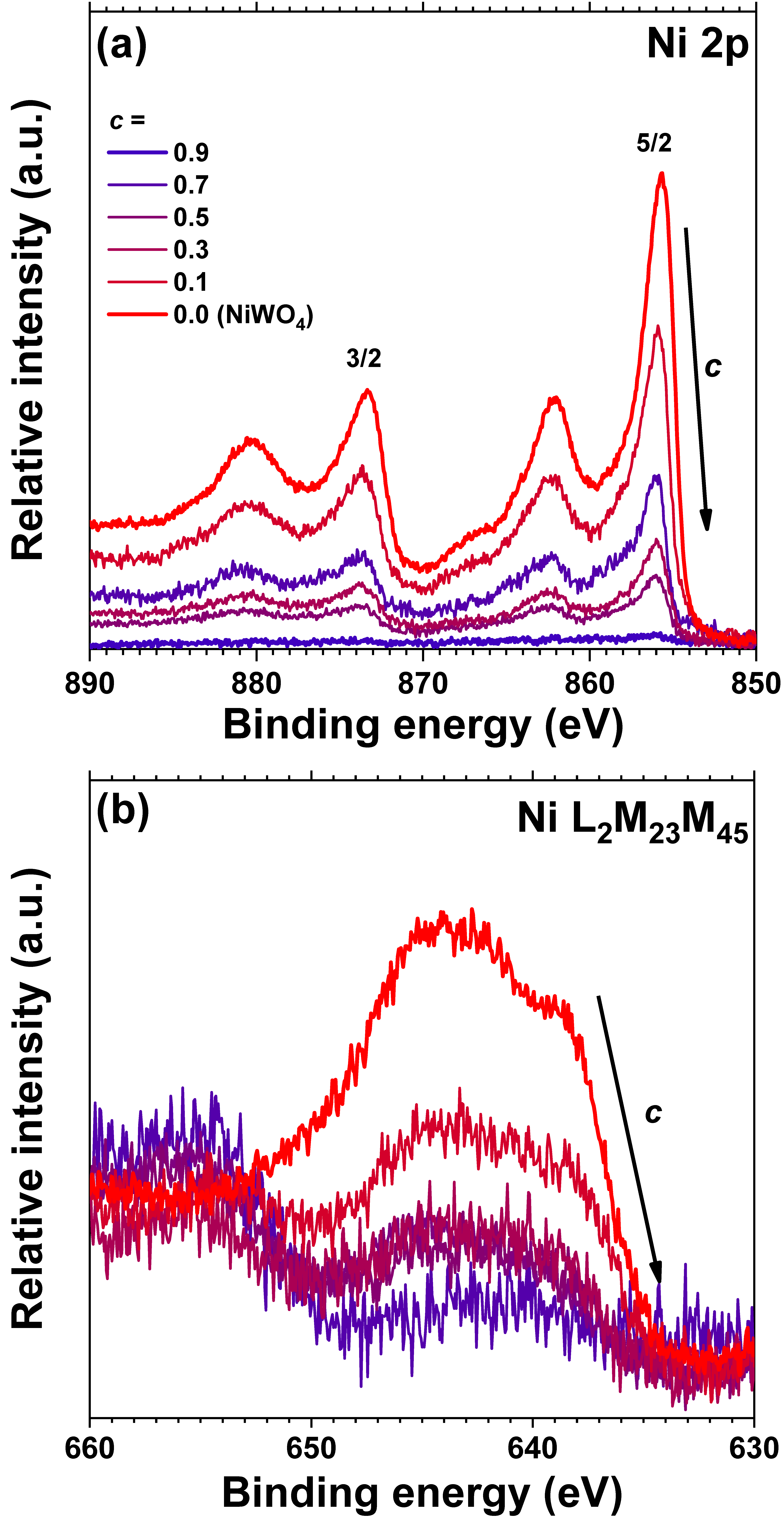}
	\caption{High-resolution X-ray photoelectron spectra of the Ni 2p region (a) and Ni $\text{L}_2\text{M}_{23}\text{M}_{45}$ Auger region (b) as recorded from \ce{Zn_cNi_{1-c}WO4} solid solutions at a detection angle of $0\degree$.}
	\label{fig2}
\end{figure}

\section{Results and discussion}
\subsection{XPS measurements}
\begin{figure}
	\centering
	\includegraphics[width=0.4\textwidth]{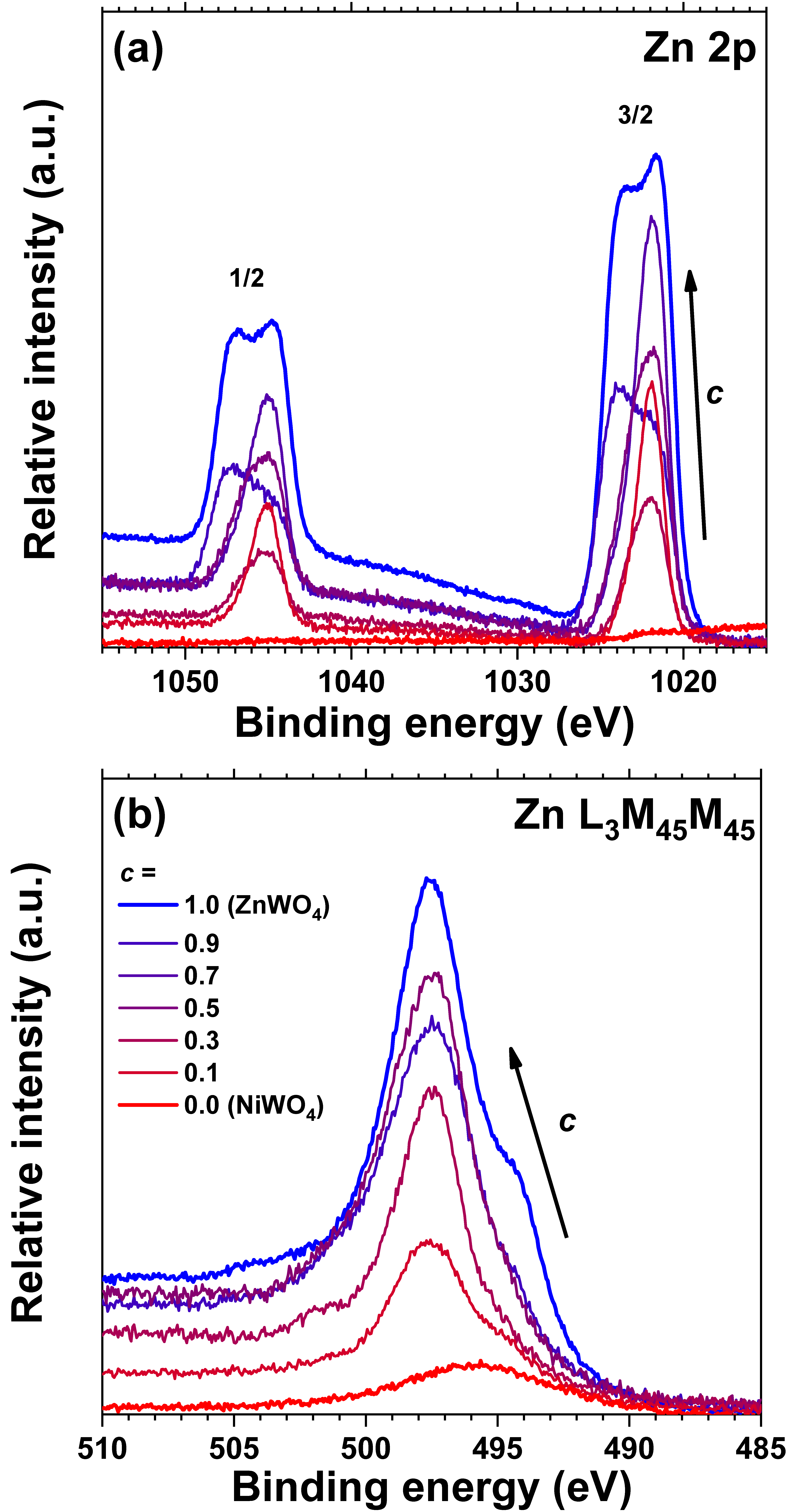}
	\caption{High-resolution X-ray photoelectron spectra of the Zn 2p region (a) and Zn $\text{L}_3\text{M}_{45}\text{M}_{45}$ Auger region (b) as recorded from \ce{Zn_cNi_{1-c}WO4} solid solutions at a detection angle of $0\degree$.}
	\label{fig3}
\end{figure}

A representative set of high-resolution X-ray photoelectron spectra for different regions is shown in Figs.~\ref{fig1}--\ref{fig4}.

\subsubsection{VB region}
The VB spectrum of \ce{NiWO4} provides the starting point for discussion of the VB region of \ce{Zn_{c}Ni_{1-c}WO4} solid solutions (see Fig.~\ref{fig1}). A first-principles linear combination of atomic orbitals (LCAO) study \cite{Kuzmin2011niwo4} has shown that in \ce{NiWO4} the VB is formed by hybridized O 2p and Ni 3d states, and the Ni 3d ($\text{t}_\text{2g}$, $\text{e}_\text{g}$, $\uparrow$) states contribute mainly to the upper part of the VB, whereas the conduction band is formed by the empty W 5d states with an admixture of the empty Ni 3d ($\text{e}_\text{g}$, $\downarrow$) states at the bottom of the band. In the VB spectrum of \ce{NiWO4} shown in Fig.~\ref{fig1}, two main spectral features may be identified: the O 2s XPS peak, and a broad VB consisting of the O 2p and Ni 3d states at lower BE. The relatively shallow Ni 3d states hybridize with O 2p states to give bonding states at the top of the main VB. The O 2p VB spectrum for \ce{Zn_cNi_{1-c}WO4} solid solutions is similar to that of the parent \ce{NiWO4}, although the overall bandwidth becomes narrower. As shown in the LCAO calculations, the VB of \ce{ZnWO4}---having largely O 2p character---is separated by a band gap of 4.6~eV from the bottom of the conduction band, which is dominated by W 5d states \cite{Kalinko2009}. The most pronounced difference between the \ce{NiWO4} and Zn-containing solid solutions is the emergence of and a gradual increase in the intensity of the Zn 3d photoelectron peak relative to that of the O 2p VB, as is to be expected from the change in composition. The VB spectra do not contain any weak in-gap structures below the Fermi level arising from states associated with oxygen deficiency \cite{Kalinko2011a}.

\subsubsection{Ni 2p and Ni LMM regions}
The Ni 2p$_{3/2}$ spectra of \ce{Zn_cNi_{1-c}WO4} solid solutions are complex because of multiplet splitting (see Fig.~\ref{fig2}a), which is clearly resolved for $c \le 0.70$; for $c \ge 0.90$, nickel is not detectable in the samples. According to the TPP-2M model \cite{tanuma1994calculations}, in \ce{NiWO4} (22 valence electrons, density 7.55~g/cm$^3$, band gap~2.2 eV), the inelastic mean free paths (IMFPs) of electrons of energy of around 600~eV (Ni 2p), 845~eV (Ni LMM), and 1450~eV (W 4f) are 13.10~\AA, 16.29~\AA, and 24.51~\AA, respectively. Thus, the Ni 2p photoelectron line carries more surface-sensitive information than the W 4f photoelectron line. As octahedral units are unlikely to be broken during sample preparation through grinding, and taking into account that the average Ni--O distance in \ce{NiWO4} is around 2~\AA, the surface in solid solutions with high $c$ values seems to be indeed depleted in Ni, and this is explained by the more covalent nature of Ni--O chemical bonds (vide infra) and, thus, higher Peierls-Nabarro barriers for dislocation motion in the planes containing \ce{[NiO6]}, so the fracture occurs preferentially along the planes containing \ce{[ZnO6]}.

High-resolution spectra of the X-ray-excited Ni LMM Auger-transition lines from selected samples are shown in Fig.~\ref{fig2}b. The signal-to-noise ratio is low. In pure \ce{NiWO4}, the Ni LMM line has a clearly resolved double-peak structure, although for intermediate compounds, intensity redistribution occurs. The subtle peak on the high-BE side is due to Zn LMM Auger transitions.

\subsubsection{Zn 2p and Zn LMM regions}
The Zn 2p$_{3/2}$ spectra of \ce{ZnWO4}---although free from multiplet splitting and other complicating effects---in all intermediate compositions consist of two components. Close to the zinc-rich end, the lower-intensity component appears as a clear shoulder on the high-BE side of the main peak (see Fig.~\ref{fig3}a). This can be explained by the formation of zinc hydroxide on the surface of the powder particles. At low temperatures, the formation of \ce{Zn(OH)2} on the surface is thermodynamically more favorable, taking into account depletion of Ni in the surface region, and a more negative standard enthalpy of formation for \ce{Zn(OH)2} than for \ce{Ni(OH)2}: $-555.9$~kJ/mol versus $-458.3$~kJ/mol, respectively \cite{rabinovich1983short}.

According to the TPP-2M model \cite{tanuma1994calculations}, in \ce{ZnWO4} (24 valence electrons, density 7.87~g/cm$^3$, band gap 4.6~eV), the IMFPs of electrons with energy of around 460~eV (Zn 2p), 985~eV (Zn LMM), and 1450~eV (W 4f) are 10.55~\AA, 18.13~\AA, and 24.33~\AA, respectively. Thus, the Zn 2p region is more surface sensitive than the W 4f region. High-resolution X-ray-excited Zn LMM Auger-transition line spectra from selected samples are shown in Fig.~\ref{fig3}b. This region is complicated by the presence of a weak W 4p$_{1/2}$ X-ray photoelectron line, whose relative contribution can be estimated from Fig.~\ref{fig3}b when no zinc is present ($c = 0$). Although the Na KLL Auger-transition line is located in this BE region, sodium was not detected in the survey.

\subsubsection{W 4f and W NNN regions}
\begin{figure}
	\centering
	\includegraphics[width=0.4\textwidth]{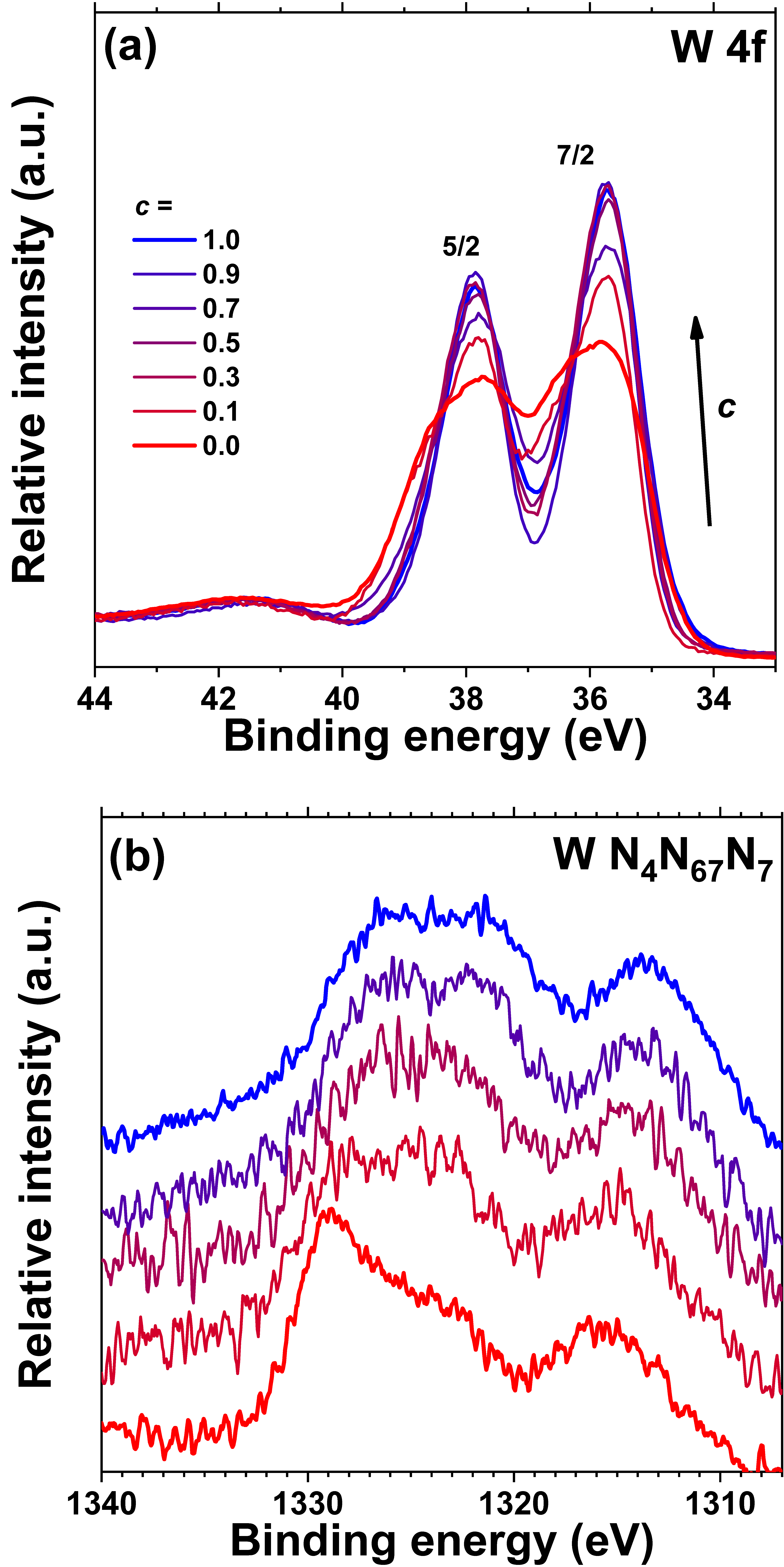}
	\caption{High-resolution X-ray photoelectron spectra of the W 4f region (a) and W $\text{N}_4\text{N}_{67}\text{N}_{7}$ Auger region (b) as recorded from \ce{Zn_cNi_{1-c}WO4} solid solutions at a detection angle of $0\degree$. The spectra of the Auger region have been shifted along the ordinate for a better overview.}
	\label{fig4}
\end{figure}

From general considerations (i.e., based on equal atomic concentrations of W and similar IMFPs for W 4f photoelectrons in \ce{NiWO4} and \ce{ZnWO4}) one would expect the relative intensity and the shape of the XPS W 4f doublet in the whole solid-solution series to be the same. However, as is apparent from Fig.~\ref{fig4}a, the overall shape of the peaks does change: from very broad in \ce{NiWO4} to much narrower in \ce{ZnWO4}. At the same time, the integral intensity of the peaks stays approximately the same. Only a few high-quality XPS studies of \ce{NiWO4} have been published so far \cite{Mancheva2007, GREEN2014248, DOUDIN201720}. The width of the W 4f doublet recorded from thin epitaxial \ce{NiWO4} films was very small \cite{DOUDIN201720}, whereas the W 4f doublet reported for powders \cite{Mancheva2007} and magnetron-sputtered thin films \cite{GREEN2014248} is characterized by considerable broadening. We assume that the distortion of \ce{[WO6]} octahedra contributes to the width of the broadened W 4f peaks (vide infra).

For chemical-state determination, usually the KE of the $\text{M}_5\text{N}_{67}\text{N}_{67}$ Auger-transition line is used (KE of around 1728~eV). As this energy was not accessible in our setup, we used the AP $\alpha(\text{4f}_{7/2}, \mathrm{N}_4\mathrm{N}_{67}\mathrm{N}_{7})$. Because of the low KE of these Auger electrons, the intensity of the line is low even in the spectrum of pure tungsten metal, and thus in solid solutions the noise level in the spectra of the W NNN Auger-transition line is very high, but it shows gradual changes from $c=0$ to $c=1$ (Fig.~\ref{fig4}b).

As noted in \cite{Schofield1993} the stoichiometry of tungstates is difficult to demonstrate. In these tungstate samples both Ni and Zn are in the +2 oxidation state, and any nonstoichiometry would depend on the presence of multiple-valency states for W, although from the W 4f X-ray photoelectron spectra there is no evidence for the latter. As mentioned above, no features associated with oxygen deficiency were observed in the VB spectra. Any nonstoichiometry or disorder within the tungstate samples would have to be a major phenomenon, but no signs of nonstoichiometry or disorder were detected in X-ray powder diffraction experiments \cite{Kalinko2011a} or in X-ray absorption spectroscopy experiments \cite{Bakradze2020}. We conclude that cation or oxygen deficiencies are not the reason for the observed width of the W 4f peak in nickel-rich samples.

\subsection{Chemical-state plots}
Fig.~\ref{fig5} shows the chemical-state plots for nickel, zinc, and tungsten in \ce{Zn_cNi_{1-c}WO4} solid solutions as a function of zinc concentration $c$ (in the plot the abscissa, BE ($c_0$), is oriented in the negative direction). Table \ref{table1} shows the maximum BE positions of the Ni 2p$_{3/2}$, Zn 2p$_{3/2}$, and W 4f$_{7/2}$ photoelectron peaks and the maximum KE positions of the Ni LMM, Zn LMM, and W NNN Auger electron peaks, as well as the AP for Ni, Zn, and W and final-state effects $\Delta R^\text{ea}$ and initial-state effects $\Delta \epsilon$ for Ni, Zn, and W (with respect to the metallic state).

\subsubsection{Ni chemical-state plot}
\begin{figure}
	\centering
	\includegraphics[width=0.4\textwidth]{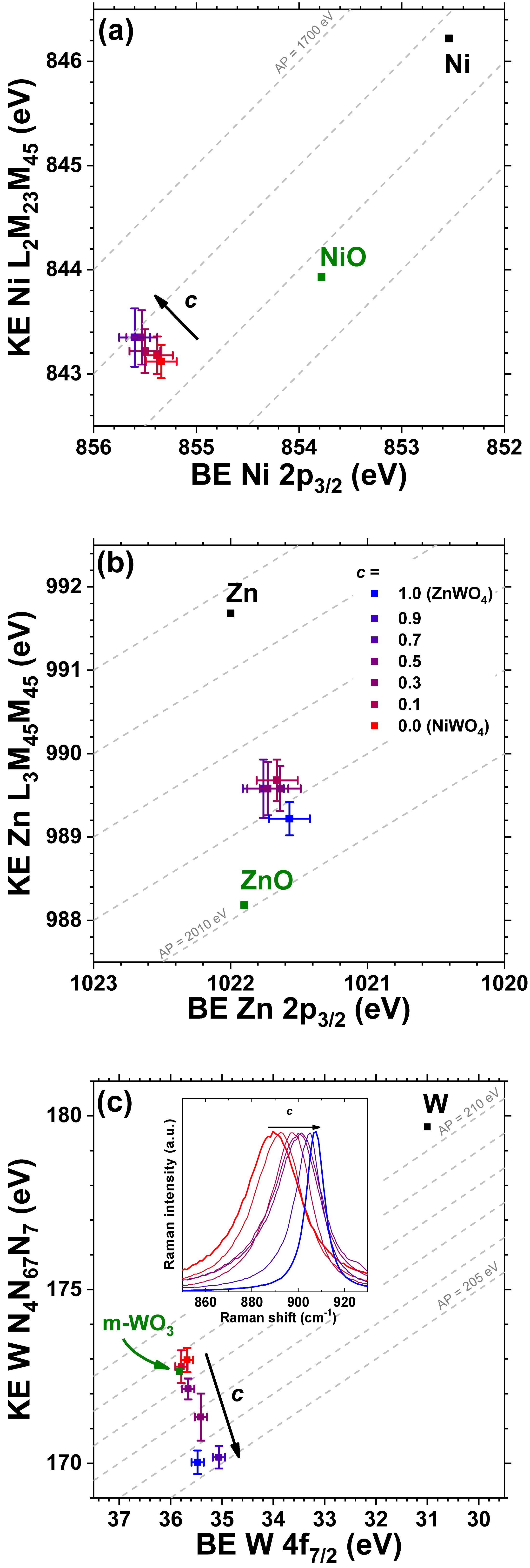}
	\caption{Chemical-state plots for Ni (a), Zn (b), and W (c) in \ce{Zn_cNi_{1-c}WO4} solid solutions. The family of dashed diagonal lines with a slope of $-1$ represents the lines of constant Auger parameter (AP) values according to Eq. (\ref{eq:eq1}). The corresponding APs for pure metals and their stablest oxides are also indicated as reported in the NIST X-ray Photoelectron Spectroscopy Database \cite{NISTXPS} and in \cite{Biesinger2012}. The arrow indicates the direction of increasing zinc concentration in the samples. The inset in (c) shows Raman scattering spectra of \ce{Zn_cNi_{1-c}WO4} solid solutions in the range of the $\mathrm{A_g}$ band due to the W--O stretching mode \cite{Bakradze2020}. See the main text for details. BE, binding energy; KE, kinetic energy.}
	\label{fig5}
\end{figure}

The BE positions of the Ni 2p$_{3/2}$ X-ray photoelectron peak showed a shift of approximately 0.22~eV between pure nickel tungstate and solid solutions (see Fig.~\ref{fig5}a). A complex multiplet structure of the Ni 2p$_{3/2}$ line and a small shift make this peak alone unsuitable for a quantitative analysis in \ce{Zn_cNi_{1-c}WO4} solid solutions. However, chemical-state characterization can be done with the AP for nickel $\alpha(\text{2p}_{3/2}, \mathrm{L}_2\mathrm{M}_{23}\mathrm{M}_{45})$ \cite{Biesinger2012}. In the chemical-state plot for nickel, with the increase of zinc concentration, the BE of the Ni 2p$_{3/2}$ line and the KE of the Ni LMM Auger-transition line slightly increase. The AP for nickel gradually increases with increasing zinc concentration, and the maximum measured change in the AP for Ni in \ce{NiWO4} and \ce{Zn_{0.70}Ni_{0.30}WO4} is approximately 0.50 eV (see Table \ref{table1}). The higher Ni 2p$_{3/2}$ BEs in solid solutions evidence the higher ionicity of Ni--O chemical bonds with increasing zinc concentration. For a core-ionized atom with an open valence shell configuration, e.g., \ce{Ni^{2+}} ($\text{[Ar]}\,3\text{d}^8$), one would expect the screening mechanism to be nonlocal: an electron transfer occurs from the nearest-neighbor ligands into the spatially extended d atomic orbitals of the 2p core-ionized \ce{Ni^{2+}} ions; thus, an AP shift should be practically independent of the chemical state and similar to that measured for the metallic state. As is apparent from Table \ref{table1}, the AP for Ni is indeed close to that for metallic Ni (1698.76~eV); however, the AP values increase gradually upon dilution of \ce{NiWO4} with zinc and are higher than those for NiO (1697.71~eV \cite{Biesinger2012}). Thus, each of these chemical states has a different AP value, i.e., relaxation effects in response to the creation of the photoelectron core hole \cite{MORETTI1998}. It is known that the AP increases with increased coordination \cite{MORETTI1998}; thus, this observation could indicate that the number of distinct neighbors of Ni increases (i.e., the \ce{[NiO6]} octahedra become less distorted with increasing zinc concentration). This qualitative finding is in agreement with a recent X-ray absorption study \cite{BAKRADZE2021} where narrowing of the radial distribution function for Ni--O atomic pairs was found.

Analysis of the initial-state and final-state effects for Ni in solid solutions indicates that at the nickel-rich end their BE and KE shifts are dominated by the initial ground electronic state effects, $\Delta \epsilon$, which increase toward the nickel-rich end and reach their maximum in \ce{NiWO4} (+0.45~eV). In the middle of the concentration range, the BE and KE shifts are determined by both the initial-state shifts and the final-state shifts, whose contributions are approximately equal: +0.06~eV versus +0.05~eV, respectively (see Table \ref{table1}).

\subsubsection{Zn chemical-state plot}
The BE positions of the Zn 2p$_{3/2}$ X-ray photoelectron peak showed a maximum shift of approximately 0.33~eV between pure \ce{ZnWO4} and solid solutions (see Fig.~\ref{fig3}b and Table~\ref{table1}). Such a small shift makes this peak unsuitable for a meaningful analysis; this observation again confirms that the use of the AP is preferred to identify zinc species. Chemical-state characterization can be done with the AP $\alpha(\text{2p}_{3/2}, \mathrm{L}_3\mathrm{M}_{45}\mathrm{M}_{45})$. The position of zinc in \ce{Zn_cNi_{1-c}WO4} solid solutions in the chemical-state plot with respect to the metallic state is essentially determined by a high negative value of the Madelung potential. At the same time, the chemical state of Zn ions in solid solutions with respect to the pure \ce{ZnWO4} state shows no systematic changes as a function of zinc concentration (see Fig.~\ref{fig5}b and Table~\ref{table1}). Thus, for \ce{Zn^{2+}} ($[\text{Ar}]\,3\text{d}^{10}$)---i.e., the core-ionized atoms with a closed valence shell configuration---the screening mechanism is local: an electron transfer occurs from the nearest-neighbor oxygen ligands into the localized p atomic orbital of the 2p core-ionized \ce{Zn^{2+}} ions and the AP shifts are practically independent of the chemical state. The AP for Zn almost does not depend on zinc concentration (see Table \ref{table1}), but never reaches the AP value of zinc in \ce{ZnO} (2010~eV). This observation is in agreement with the tendency of the AP to increase with increased coordination \cite{MORETTI1998}: in \ce{ZnWO4}, Zn has an octahedral coordination, whereas in \ce{ZnO} the coordination is tetrahedral.

\subsubsection{W chemical-state plot}
Chemical-state characterization of tungsten was done with the AP $\alpha(\text{4f}_{7/2}, \text{N}_4\text{N}_{67}\text{N}_{7})$. The chemical state of W ions in solid solutions changes considerably as a function of zinc concentration (see Fig.~\ref{fig5}c and Table~\ref{table1}). A shift of the W 4f$_{7/2}$ BE with increasing zinc concentration evidences a more covalent character of the W--O chemical bonds with increasing zinc concentration. Each of these chemical states has a different AP value involving variable extra-atomic polarization charge toward the spatially extended valence orbitals of the 4f core-ionized \ce{W^{6+}}.

\begin{sidewaystable}
	\small
	\centering
	\caption{Zinc content in solid solutions, $c$, binding energy (BE) positions, $E_\text{b}$, of Ni 2p$_{3/2}$, Zn 2p$_{3/2}$, and W 4f$_{7/2}$ photoelectron peaks, kinetic energy (KE) positions, $E_\text{k}$, of Ni $\mathrm{L}_2\mathrm{M}_{23}\mathrm{M}_{45}$ (Ni LMM), Zn $\mathrm{L}_3\mathrm{M}_{45}\mathrm{M}_{45}$ (Zn LMM), and W $\mathrm{N}_4\mathrm{N}_{67}\mathrm{N}_{7}$ (W NNN) Auger-transition peaks, Auger parameters, $\alpha$, for Ni, Zn, and W, and final-state effects, $\Delta R^\text{ea}$, and initial-state effects, $\Delta \epsilon$, for Ni, Zn, and W (with respect to the metallic state). All energy data are given in electronvolts.}\label{table1}
	\begin{tabular}{@{}llllllllllllllll@{}}
		\toprule
		$c$ & Ni 2p$_{3/2}$ BE & Ni $\text{LMM}$ KE & Zn 2p$_{3/2}$ BE & Zn $\text{LMM}$ KE & W 4f$_{7/2}$ BE & W $\text{NNN}$ KE & $\alpha_\text{Ni}$ & $\alpha_\text{Zn}$ & $\alpha_\text{W}$ & $\Delta R^\text{ea}_\text{Ni}$ & $\Delta R^\text{ea}_\text{Zn}$ & $\Delta R^\text{ea}_\text{W}$ & $\Delta \epsilon_\text{Ni}$ & $\Delta \epsilon_\text{Zn}$ & $\Delta \epsilon_\text{W}$ \\
		\midrule
		1.0 & --     & --     & 1021.57 & 989.22 & 35.47 & 170.03 & --      & 2010.79 & 205.50 &--    & $-1.45$ & $-2.59$ & --    & 1.88 & $-1.88$ \\
		0.9 & --     & --     & 1021.76 & 989.58 & 35.06 & 170.17 & --      & 2011.34 & 205.23 &--    & $-1.17$ & $-2.73$ & --    & 1.41 & $-1.34$ \\
		0.7 & 855.60  & 843.35 & 1021.73 & 989.58 & 35.66 & 172.14 & 1698.95 & 2011.31 & 207.80 &+0.10 & $-1.19$ & $-1.97$ & $-0.06$ & 1.46 & $-1.63$ \\
		0.5 & 855.53 & 843.35 & 1021.64 & 989.56 & 35.41 & 171.33 & 1698.88 & 2011.20 & 206.74 &+0.06 & $-1.23$ & $-1.44$ & +0.05 & 1.57 & $-4.03$ \\
		0.3 & 855.50  & 843.22 & 1021.66 & 989.68 & --    & --     & 1698.72 & 2011.34 & --     &$-0.02$ & $-1.17$ & --    & +0.16 & 1.51 & --    \\
		0.1 & 855.38 & 843.18 & --      & --     & 35.79 & 172.78 & 1698.56 & --      & 208.57 &$-0.10$ & --    & $-1.06$ & +0.36 & --   & $-3.74$ \\
		0.0 & 855.34 & 843.12 & --      & --     & 35.68 & 172.97 & 1698.46 & --      & 208.65 &$-0.15$ & --    & $-1.02$ & +0.45 & --   & $-3.67$ \\
		\bottomrule
	\end{tabular}
\end{sidewaystable}

Similarly as in the case of Zn, i.e., for a core-ionized atom with a closed valence shell configuration, for $\text{W}^{6+}$ ($\text{[Xe]}\,4\text{f}^{14}$) one would expect the screening mechanism of the 4f ionized W to be local. However, the analysis of the initial-state effects, $\Delta \epsilon$, and final-state effects, $\Delta R^\text{ea}$, for W in solid solutions indicates that for the nickel-rich end, the BE and KE shifts are dominated by the initial ground electronic state effects, which are more negative in \ce{NiWO4} than in \ce{ZnWO4}: $-3.67$~eV versus $-1.88$~eV, respectively (see Table \ref{table1}). At the zinc-rich end, their BE and KE shifts are dominated by the final-state shifts, which are more negative in \ce{ZnWO4} than in \ce{NiWO4}: $-2.59$~eV versus $-1.02$~eV, respectively. The AP depends on the electronic polarizability of the nearest-neighbor ligands, and it can be shown \cite{MORETTI1998, MORETTI2013} that the final-state relaxation energy decreases when the number of ligands or their polarizability decreases. The polarization of the ligands may be seen as an electron transfer from the nearest-neighbor ligands toward the spatially extended valence orbitals of the 4f core-ionized W: the shared W--O electrons are polarized back toward O, resulting in the lower measured BEs. This interpretation is backed up by extended X-ray absorption fine structure spectroscopy results obtained for \ce{Zn_cNi_{1-c}WO4} solid solutions \cite{Bakradze2020} and the Raman data on the $\text{A}_\text{g}$ band shift in \ce{Zn_cNi_{1-c}WO4} solid solutions (see the inset in Fig.~\ref{fig5}c and \cite{Bakradze2020}). Although the formal coordination number of W in \ce{Zn_cNi_{1-c}WO4} solid solutions (coordination number of 6) does not change, the large AP shift is attributed to the changes of W--O interatomic distances as a function of $c$. In pure \ce{NiWO4}, the closest oxygen atoms split into two groups of 4 $\times$ 1.84~\AA\ and 2$\times$ 2.12~\AA, whereas in pure \ce{ZnWO4}, the closest oxygen atoms split into three groups of 2 $\times$ 1.80~\AA, 2 $\times$ 1.90~\AA, and 2 $\times$ 2.14~\AA, thus changing the local static distortion of \ce{[WO6]}. In solid solutions, the W--O distances gradually take intermediate values with increasing zinc concentration; see \cite{Bakradze2020} for details. In both \ce{NiWO4} and \ce{ZnWO4}, the \ce{[WO6]} octahedra are distorted because of the second-order Jahn-Teller effect \cite{KUNZ1995}. Although the temperature-induced effects in the \ce{[WO6]} octahedra are more pronounced in \ce{ZnWO4}, the W--O bonds are stiffer in \ce{ZnWO4}, as is evidenced by (i) the larger effective bond-stretching force constants \cite{BAKRADZE2021} and (ii) the higher frequency of the W--O stretching $\mathrm{A_g}$ mode determined by Raman spectroscopy reported in \cite{Bakradze2020}. This distortion and weaker W--O bond strength in \ce{NiWO4} are the reasons for the increased width of the W 4f X-ray photoelectron lines.

\section{Conclusions}
Spectra of Ni 2p, Zn 2p, and W 4f photoelectron lines and Ni LMM, Zn LMM, and W NNN Auger-transition lines were recorded from the whole range of \ce{NiWO4}-\ce{ZnWO4} ternary-oxide solid solutions. Upon dilution with zinc ions, the spectra show pronounced changes in the shape and main line positions, which are reflected in the chemical-state  plots for Ni, Zn, and W ions. The chemical-state plots for Ni, Zn, and W ions indicate that the AP slightly increases for Ni (+0.49~eV) and substantially decreases for W ($-3.15$~eV) with increasing zinc concentration. These changes are explained by the concurrent changes in the first coordination spheres of both Ni and W. The changes are accompanied by an increase and lowering of the electronic polarizability of the first coordination sphere around Ni and W atoms, respectively. It is concluded that with increasing zinc concentration, the ionicity of Ni--O chemical bonds increases, the ionicity of W--O bonds decreases (i.e., W--O bonds become more covalent), and the ionicity of Zn--O chemical bonds does not change. Final-state effects contribute unequally within the solid-solution series: the external relaxation effects for Ni are found to slightly increase with increasing zinc concentration, in agreement with the shortening of Ni--O nearest-neighbor bonds, whereas for W the external relaxation effects are found to decrease with increasing zinc concentration, in agreement with the shortening of W--O nearest-neighbor bonds. The data indicate that upon zinc substitution in \ce{Zn_cNi_{1-c}WO4} solid solutions, the environment around W modifies the ground-state valence charge of W and, therefore, the ability of the valence charge to screen the final-state hole. Although a careful in-depth analysis of XPS data obtained with a laboratory-based XPS system can give only qualitative information on the changes in the first coordination spheres of each metal ion, this information is useful, because otherwise it is accessible only by synchrotron-based techniques (such as X-ray absorption spectroscopy).

\section*{Declaration of competing interest}

The authors declare that they have no known competing financial interests or personal relationships that could have appeared to influence the work reported in this article.

\section*{Acknowledgments}
G. Bakradze acknowledges the financial support provided by the State Education Development Agency for project no. 1.1.1.2/VIAA/3/19/444 (agreement no. 1.1.1.2/16/I/001) realized at the Institute of Solid State Physics, University of Latvia. A. Kalinko and A. Kuzmin  thank the  Latvian Council of Science for support (project no. lzp-2019/1-0071).
The Institute of Solid State Physics, University of Latvia, as a center of excellence has received funding from the European Union's Horizon 2020 Framework Programme H2020-WIDESPREAD-01-2016-2017-TeamingPhase2 under grant agreement no. 739508, project CAMART$^2$.

\end{document}